\documentclass[twocolumn, prl, showpacs]{revtex4}

\usepackage{amssymb}
\usepackage{graphicx}
\usepackage{dcolumn}
\usepackage{bm}
\usepackage{amsmath}

\begin{document}
\title{Quantum Spinon Oscillations}

\author{Zi Cai$^1$, Lei Wang$^2$ , X. C. Xie$^{3,4,2}$, U. Schollw\"{o}ck$^5$, X.R. Wang$^6$, M. Di Ventra$^1$
, Yupeng Wang$^2$}

\affiliation{$^{1}$Department of Physics, University of California,
San Diego, California 92093, USA}

\affiliation{$^{2}$ Beijing National Laboratory for Condensed Matter
Physics, Institute of Physics, Chinese Academy of Sciences, Beijing
100080, P. R. China}

\affiliation{$^{3}$International Center for Quantum Materials,
Peking University, Beijing 100871, China}

\affiliation{$^{4}$Department of Physics, Oklahoma State University,
Stillwater, Oklahoma 74078, USA}

\affiliation{$^{5}$Fakult\"{a}t f\"{u}r Physik,
Ludwig-Maximilians-Universit\"{a}t M\"{u}nchen, D-80798 M\"{u}nchen,
Germany}

\affiliation{$^{6}$Physics Department, The Hong Kong University of
Science and Technology, Clear Water Bay, Hong Kong SAR, China}

\begin{abstract}
The full quantum dynamics of a spinon under external magnetic fields
is investigated by using the time-evolving block decimation (TEBD)
method within the microcanonical picture of transport. We show that the center of the spinon oscillates back and
forth in the absence of dissipation. The quantum many-body behavior
can be understood in a single-particle picture of transport and
Bloch oscillations, where quantum fluctuations induce finite life
times. Transport, oscillations and lifetimes can be tuned to some
degree separately by external fields. Other nontrivial dynamics such
as resonance as well as chaos have also been discussed.
\end{abstract}
 \maketitle

Controlling the time-evolution of quantum states and manipulating
the dynamics of quantum matter have attracted considerable
theoretical and experimental attention in recent years due to their
relevance to fundamental physics as well as potential applications
in  information storage, encoding and processing. Up to now, most
efforts in this field are based on solid state systems. Recently,
the progress in ultracold atomic gases in optical lattices not only
provides an ideal platform for simulating quantum many-body models
in condensed matter physics, but also paves the way for quantum
manipulation. Due to the unique properties of the ultracold atomic
gases in optical lattices, such as the extremely low dissipation
rates and long coherence times, we can explore the dynamics of
quantum many-body systems in a clean and dissipation-less
environment. This enables us to manipulate the quantum many-body
state with an unprecedented degree of precision, and study new physical effects\cite{Bloch}. An early breakthrough in the field was the
first direct observation  of  Bloch oscillations (BO) in a tilted
optical lattice\cite{Dahan,Wilkinson}. In solid state systems,
scattering by impurities would result in strong damping of Bloch
oscillations, while in an optical lattice the perfect optical
crystals are free from any imperfections, and what is more, the long
coherence times and high tunability of optical lattices not only
enable us to observe BO directly, but also to manipulate the
dynamics of the quantum particles  via external fields.

Compared with the dynamics of the quantum particles, it may be more
interesting to explore and manipulate the quasi-particles, which
emerge as  elementary excitations in quantum systems consisting of a
collection of interacting particles, and may behave differently
compared to  the original particles. The most interesting example of this
is known as fractionalization: the particles are effectively split
into smaller constituent quasi-particles, which only carry a
fraction of quantum numbers.

In this paper, we study the quantum
dynamics of one of the most known quasi-particles: the spinon, which
is an excitation separating two degenerate states of opposite
magnetization in a quantum spin chain and usually carries a
fractionalized quantum number (spin-$\frac 12$). Using the
time-evolving block decimation (TEBD) method\cite{Vidal,Schollwock} within the microcanonical
picture of transport~\cite{Di-Ventra2004,Di-Ventra2005},
we study the time evolution from a quantum many-body initial state
with a spinon, and show how to manipulate its dynamics via
external fields.  In particular, we find that this quantum many-body
problem can be interpreted in the picture of a quantum quasiparticle
at the single-particle level, which shows controllable linear
transport and (Bloch) oscillation behavior depending on
perpendicular external fields, where the field along the
magnetization direction controls the Bloch oscillation physics and
the transverse field controls the linear velocity. The broadening of
the quasiparticle is also tunable by the field strength. We also
find a quantum resonance behavior of the spinon under a
periodically driven external field. This may pave the way to the
controlled manipulation of quasiparticles, also in view of
potential applications in condensed matter and ultracold atom
physics.

Our Hamiltonian is a finite 1D transverse Ising model with an additional
magnetic field along the $z$-direction:
\begin{equation}
 H= \sum_iJ\sigma^z_i \sigma^z_{i+1}+g\sigma^x_i+h\sigma^z_i\\
\label{onsite}
\end{equation}
with a ferromagnetic coupling $J=-1$, Pauli matrices
${\sigma_i^{x,z}}$ on sites $i$, and a transverse field along the
$x$-direction that breaks magnetization conservation. Without the
magnetic field $h\sigma^z_i$, the transverse Ising model can be
solved exactly by a Jordan-Wigner transformation and is known as a
classic paradigm of a quantum phase transition\cite{Sachdev}.

\begin{figure}[htb]
\includegraphics[width=8.5cm]
{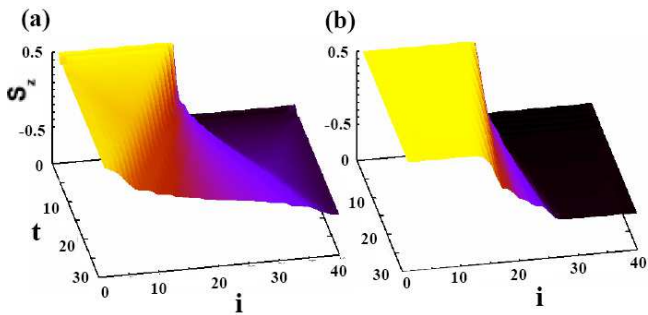} \caption{The time evolution of the spinon for $h=0$ and
(a) $g=0.1$, (b) $g=0.3$.} \label{COM}
\end{figure}

Recently, the transverse Ising model  has been realized
experimentally in cobalt niobate and a $E_8$ symmetry has
been observed in the vicinity of the quantum critical point of this
model\cite{Coldea,Zamolodchikov}. In the presence of a magnetic
field along the $z$-direction $h$, Eq.(\ref{onsite}) can no longer
be solved exactly by Jordan-Wigner transformation, because it would
induce a nonlocal term.  In this paper, $h$ is a constant or a
time-dependent field, and in both case, it will lead to nontrivial
dynamics of the spinon.

\begin{figure}[htb]
\includegraphics[width=7.0cm]
{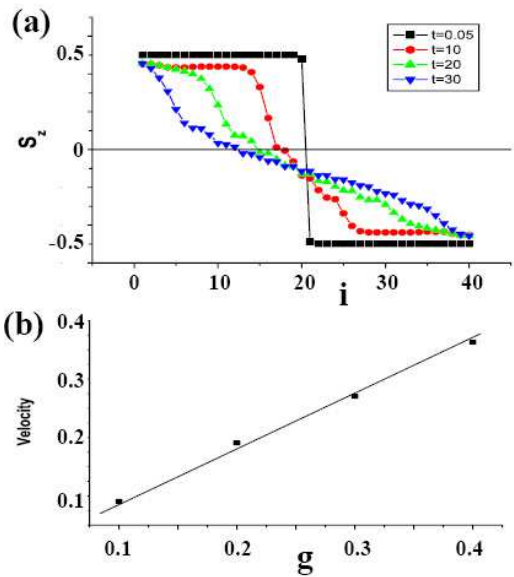} \caption{ (a) Magnetization $S^z_i$ at different times
(black squares: $t=0.05$; red dots: $t=10$; green upward triangles:
$t=20$; blue downward triangles: $t=30$) for $h=0$, $g=0.3$. (b) The
dependence of the spinon velocity on $g$.} \label{data1}
\end{figure}

\begin{figure}[htb]
\includegraphics[width=8.5cm]
{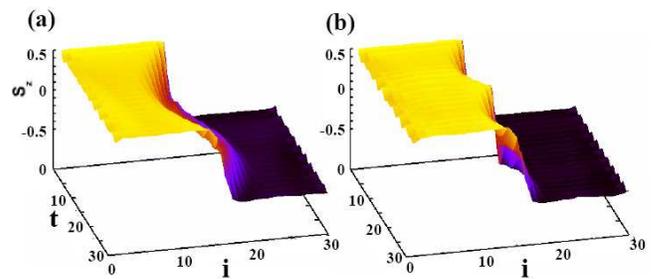} \caption{The time evolution of a spinon for $g=0.3$ and
(a) $h=0.1$, (b) $h=0.2$.} \label{DWW}
\end{figure}

Our many-body initial state is prepared as follows:  a spinon with a
finite velocity is located at the center of lattice. To realize this
initial state, we can choose its wavefunction as
$|\varphi\rangle_{t=0}=\frac{1}{N_0}\sum_i A_ie^{{\rm i}k_0
i}|i\rangle$, where $N_0$ is the normalization constant and
$|i\rangle$ denotes a perfect static spinon located between site $i$
and $i+1$:
$|i\rangle=|\uparrow\rangle_1\cdots|\uparrow\rangle_i|\downarrow\rangle_{i+1}\cdots|\downarrow\rangle_L$,
where $L$ is the length of the lattice. If we choose $A_i=1$, the
initial state is a perfect Bloch wave function with definite wave
vector $k_0$, and the spinon is totally delocalized over the length of the lattice. To ensure
that the spinon is initially located at the center of the lattice,
we then choose a Gaussian $A_i=e^{-2(i-L/2)^2}$. Given the exponentially
fast decay of $A_i$ away from the center of the lattice, we can just
keep $A_{L/2}$, $A_{L/2+1}$, $A_{L/2-1}$ and make all other $A_i=0$
for constructing the initial wave function. In our calculation we
furthermore choose $k_0=\pi/2$. We also define the average position
of the spinon $P$ as the position where $S_z=0$.

\begin{figure}[htb]
\includegraphics[width=7.0cm]
{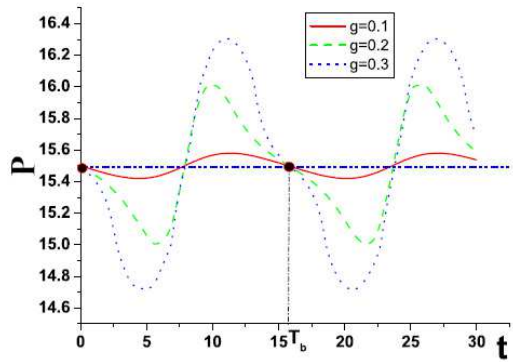} \caption{The time evolution of the position of the
spinon for $h=0.2$ and different $g$.} \label{data2}
\end{figure}

As certain excitations can be modeled as quasiparticles, i.e., having similar properties to real particles, we may conjecture that this is the case here and compare the spinon to a real particle,
where the field coupling to $\sigma^x$ lead to the motion
of the particle and where the field coupling to $\sigma_z$ acts as a
constant electric field along the axis in the particle picture, as the
Zeeman energy induced by $h\sigma_z$ is proportional to the displacement
of the spinon from its original position. To verify this picture
quantitatively, we use TEBD within the microcanonical picture of transport to calculate the time evolution of the
spinon from our initial state $|\varphi\rangle_{t=0}$ which is not an eigenstate of the Hamiltonian. In the course
of the real time evolution we use the length of our lattice $L=30$ and
take the truncation dimension $\chi=80$ and time step
$\Delta\tau=0.05$. The convergence is checked by taking larger
$\chi$. Times are measured in inverse hopping strengths $|J|^{-1}$ throughout the paper.

\begin{figure}[htb]
\includegraphics[width=7.0cm]
{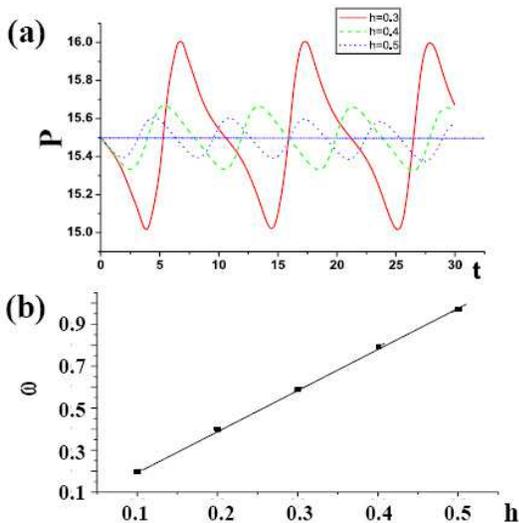} \caption{(a) The time evolution of the position of the
spinon for $g=0.3$ and different $h$; (b) the linear relation
between the frequency of BO and $h$.} \label{data3}
\end{figure}

First we focus on the case $h=0$; the real-time evolution from our
initial state $|\varphi\rangle_{t=0}$ for different parameters $g$
is shown in Fig.~\ref{COM}: (a) $g=0.1$; (b) $g=0.2$. We find that
the spinon propagates along the $-x$-direction with a velocity
proportional to $g$ as shown in Fig.~\ref{data1} (a). The linear
relation between the velocity and $g$ has been numerically verified
in Fig.~\ref {data1}(b). The quantum fluctuations increase the width
of the spinon during the time evolution, giving a lifetime to the
quasiparticle. The quantum fluctuations make the quantum spinon
different from its classical counterpart: the magnetic domain wall,
which corresponds to a soliton solution of the nonlinear
Landau-Lifshitz-Gilbert (LLG) that can preserve its shape during the
propagation. Below we will see that all these results are modified
when we introduce the magnetic field along the $z$-direction.

Next, we apply a constant magnetic field $h$ along the
$z$-direction. By interpreting the dynamics of the spinon in terms of a
real particle in a constant external field suggests the observation
of Bloch oscillations. First we fix $g=0.3$ and investigate the
dependence of the spinon dynamics on $h$. The time evolution of the
spinon under different magnetic fields $h$ is shown in
Fig.~\ref{DWW}(a) $h=0.1$; (b) $h=0.2$. For Bloch oscillations (BO)
the frequency of the BO is only dependent on the strength of the
external constant field, which can also be verified via our
numerical result, as shown in Fig.~\ref{data2}, where we fix
$h=0.2$. We can find that for different $g$, the period of the BO is
the same ($T_b=15.5$), while the amplitude of BO $A$ is determined
by $g$. If we fix $g=0.3$, the time evolution of the spinon for
different $h$ is shown in Fig.~\ref{data3}(a). Following the
interpretation in terms of a Bloch oscillation, we expect that the
oscillation frequency to be proportional to the strength of the
'external field' $h$, which can be verified numerically
(Fig.~\ref{data3}(b)).

\begin{figure}[htb]
\includegraphics[width=7.5cm]
{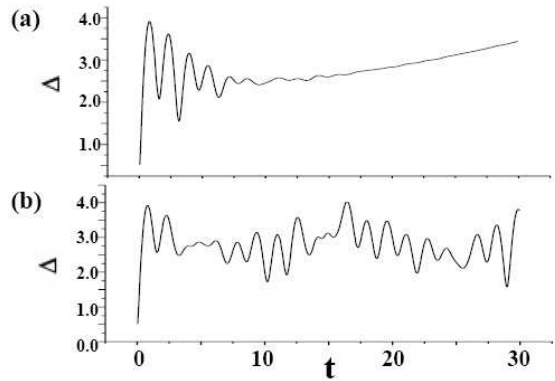} \caption{The time evolution of the width $\Delta$ of the
spinon for $g=0.1$ and (a) $h=0.0$; (b) $h=0.2$.} \label{width}
\end{figure}

Next, we study the time evolution of width of the spinon. The width
of the spinon $\Delta$ can be defined similarly to the definition of
the width of a wavepacket. We introduce $n_i=0.5-|{S^z_i}|$ and
 $ \Delta=\sqrt{\langle n\rangle^2-\langle n^2\rangle}$,
, where $\langle O\rangle=\frac{\sum_i iO_i}{\sum_i O_i}$. The
result is shown in Fig.~\ref{width}. Without the magnetic field
along the $z$-direction, $h=0$, we observe (Fig.~\ref{width} (a))
that the width oscillates strongly at first whereas after some
relaxation time, it grows linearly in time, which means the spinon
would become wider and wider in the process of propagation of the
spinon. The situation is different for $h\neq 0$, as shown in
Fig.\ref{width} (b), where after some short-time behavior the width
of the spinon oscillates around an average value instead of
diverging,  which corresponds to a solitonic breather mode: not only
is the position of the spinon confined to oscillate around the
center of the lattice by the magnetic field, but its width
oscillates as well around an average value.

\begin{figure}[htb]
\includegraphics[width=8.0cm]
{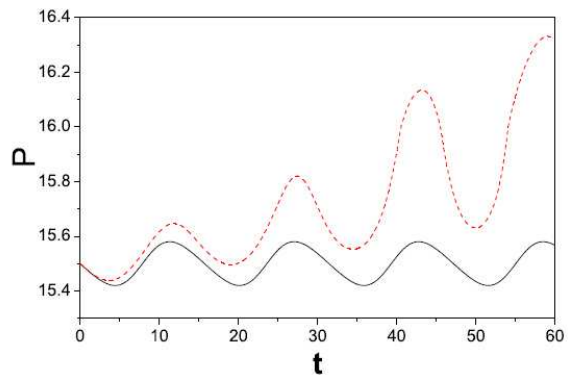} \caption{The time evolution of the position of the
spinon for $g=0.1$,  $h=0.2$ (solid line) and
$h=0.2+0.05\cos(\omega_0t)$ .} \label{res}
\end{figure}

The dynamics of the spinon is even more interesting when the system
is driven by a time-dependent magnetic field $h(t)$. In our case, we choose a periodic driving magnetic
field: $h(t)=h_0+\delta\cos(\omega t)$. Classically, it is known
that if the frequency of the driving force can match the intrinsic
frequency of the system, the system can accumulate  vibrational
energy and even a small periodic driving force can produce large
amplitude oscillations, in another words, a resonance occurs. This
phenomenon can also be observed in our quantum system, where the
intrinsic frequency corresponds to the frequency of Bloch
oscillations of the spinon. As shown in Fig.\ref{res}, the amplitude
of the oscillation of spinon is drastically enhanced when the
frequency of the driving potential is comparable to $\omega_0$, the frequency of
BO corresponding to $h=h_0$. Highly nontrivial dynamics may emerge
when the frequencies of the external driving force and the Bloch
oscillation become incommensurate, which could induce chaotic
dynamics of the spinon.

Now we discuss several possible trapped ultracold atoms systems that
could realize our Hamiltonian (\ref{onsite}) experimentally. Likely
candidates are Rydberg atoms \cite{Gallagher}, which have been
already used in quantum simulations and manipulation and have
attracted lot of attention in recent years \cite{Raitzsch,Johnson}.
Albeit highly excited, the lifetime of Rydberg atoms can reach even
hundreds of microseconds, which is much longer than the typical
lifetime of the first excited state of alkali metal atoms. Two
possible states of a Rydberg atom (excited state and ground state)
on each site reduce the problem to that of a pseudo-spin system, and
the strongly dipole-dipole interactions contribute to the
interaction between the pseudo-spins ($J\sigma^z_i\sigma^z_j$), and
the external magnetic fields $\sigma_i^x$ and $\sigma_i^z$ can be
realized by a laser with single atom Rabi frequency $\Omega_0$ and
detuning $\Delta$ respectively. Notice that the  dipole-dipole
interaction drops off  as $r^{-3}$, meaning that the next-nearest
neighbor interaction will still be appreciable, however it is much
weaker than the NN neighbor interactions. Therefore, we do not
expect that this changes the physics qualitatively for a
ferromagnetic interaction.  An alternative way would be provided by
ultracold dipolar atoms or molecules with two internal states  in
optical lattices in the hardcore limit, with an externally driven
intra-species transition ($g$-field) and a Zeeman term\cite{Rabl}.
The recent experimental progress in observing ultracold atoms at the
single atom level (or even single spin
level)\cite{Bakr,Sherson,Weitenberg} would be helpful to detect the
dynamics of the single quasi-particles (spinons). The classic
counterpart, a magnetic domain wall, has been realized
experimentally as a super cold atom thermometer\cite{Weld}.
Considering the rapid development of the field, we expect
Hamiltonian (\ref{onsite}) and the nontrivial spinon dynamics
proposed in this paper should be accessible experimentally in
various setups soon.

Our results could also be illuminating for condensed matter physics,
especially for the domain wall motion in magnetic nanowires.
Classically, the magnetization dynamics of magnetic systems is
governed by LLG equation\cite{Gilbert},
where the spin is considered as a classic vector and its dynamics is
reduced to a classic nonlinear equation. However, with advances in
the miniaturization of magnetic structures one has to anticipate
that the classical and phenomenological LLG equation will eventually
be inadequate, necessitating a full microscopic quantum description
of the magnetization dynamics. Our result actually paves the
way  to study the motion of the domain wall in the quantum situation. While
dissipation would be very weak in a quantum optical implementation
of our model, it would be more relevant in a solid. For the classic
domain wall, the dissipation is known to play a key role in determining its
velocity\cite{Wang1}. Under the assumption of a
memory-free bath, dissipation could be modeled in the framework of a
Lindblad quantum master equation or a stochastic Sch\"rodinger equation~\cite{Di-VentraSSE}; the numerical method used here can
be extended to simulate this model, both in an matrix product
operator (MPO)\cite{Verstraete,Zwolak} or quantum jump
approach \cite{Daley2}. Depending on the detailed nature of the physical
realization (and hence the bath), much richer physics could emerge
due to the interplay between interactions and dissipation making it an interesting topic for future studies.

\section*{Acknowledgment}
The authors are grateful to Congjun Wu for helpful discussion. ZC is
supported by AROW911NF0810291. M.D. acknowledges partial support
from the National Science Foundation (DMR-0802830). The work is also
supported in part by NSF-China, MOST-China and US-DOE-FG02-04ER46124
(XCX).

\end{document}